\documentclass[a4paper,11pt]{article}
\pdfoutput=1 % if your are submitting a pdflatex (i.e. if you have
\usepackage{jinstpub} % for details on the use of the package, please
\title{\boldmath Study of  cryogenic photomultiplier tubes for the future two-phase 
cryogenic avalanche detector}

\author[a,b]{A. Bondar,}
\author[a,b]{A. Buzulutskov,}
\author[b]{A. Dolgov,}
\author[a,b]{E. Frolov,}
\author[a,b]{V. Nosov,}
\author[a,b,1]{L. Shekhtman,\note{Corresponding author}}
\author[a,b]{A. Sokolov}

\affiliation[a]{Budker Institute of Nuclear Physics, SB RAS,\\630090 Novosibirsk, Russia}
\affiliation[b]{Novosibirsk State University,\\630090 Novosibirsk, Russia}

\emailAdd{L.I.Shekhtman@inp.nsk.su}

\abstract{We report the results of a characterization study of several types of cryogenic photo-multipliers manufactured 
by Hamamatsu Photonics and intended for operation in liquid Ar conditions, namely: compact 2-inch R6041-506MOD tubes, 
3-inch R11065-10 and R11065-MOD tubes for operation 
in liquid Ar and 3-inch R11410-20 tubes originally designed for operation in liquid Xe. 
These types of PMT are proposed for installation 
into the future two-phase cryogenic avalanche detector that is developed in the Laboratory of Cosmology and 
Particle Physics of the Novosibirsk State University jointly with the Budker Institute of Nuclear Physics.  
Eight R11065 PMTs and seven R11410-20 
tubes were tested and all demonstrated excellent performance in liquid Ar in terms of gain and relative single electron 
efficiency. All 3-inch PMTs showed a maximal gain in liquid Ar above 5x10$^6$ and relative single electron efficiency 
higher than 95\%.  Compact R6041-506MOD tubes have dynode  system different from that of the 3-inch photomultipliers 
and thus their single electron energy resolution 
and relative efficiency is much worse than that of 3-inch tubes. From 21 2-inch PMTs only 12 tubes were selected 
with acceptable, i.e. higher than 75\%, relative single electron efficiency  and the maximal gain higher than 5x10$^6$. However, 
these PMTs are very attractive because they are the only compact type of tubes that can operate in liquid Ar. }


\proceeding{Instrumentation for Colliding Beam Physics, INSTR17\\
  27 February - 3 March, 2017\\
  Novosibirsk, Russia}

\begin{document}
\maketitle
\flushbottom

\section{Introduction}
\label{sec:intro}

Two-phase Cryogenic Avalanche Detector (CRAD) is proposed and being developed in the Laboratory of Cosmology and Particle Physics 
of the Novosibirsk State University jointly with the Budker Institute of Nuclear Physics ~\cite{CRAD1,CRAD2,CRAD3,CRAD4,CRAD5}. 
The main purpose of this detector 
is to search for weakly interacting massive particles (WIMP)~\cite{DarkM1}  and coherent neutrino-nucleus scattering 
~\cite{cohneut1, cohneut2} by detecting signals 
from nuclear recoils~\cite{Recoils}. The energy of recoil nuclei from WIMP is predicted not to exceed several tens of keV with the most 
interesting range below 8 keV that corresponds to the WIMP mass below 10 GeV. For coherent neutrino-nucleus scattering 
the recoil energy is even lower, below 0.5 keV. Thus the CRAD sensitivity that can allow detection of single primary 
electrons becomes a factor of crucial importance.

\begin{figure}[htbp]
\centering
\includegraphics[width=1.0\textwidth,viewport=1 1 700 500,clip]{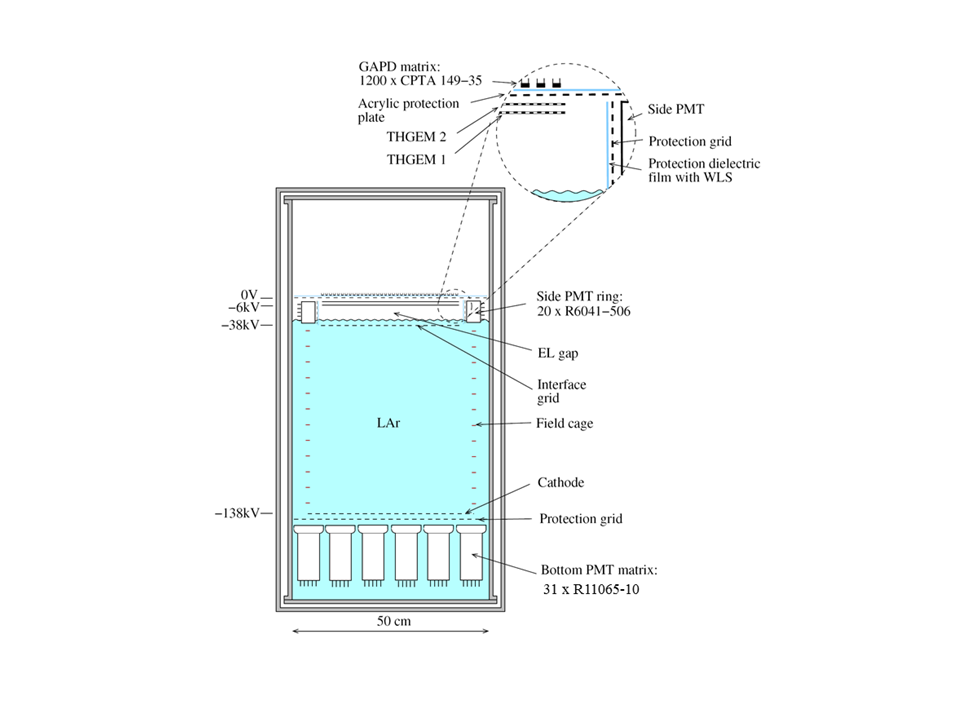}
\caption{Design of two-phase CRAD in Ar with optical readout by a combined GEM/GAPD multiplier.}
\label{fig:CRAD}
\end{figure}

The design of proposed CRAD is shown schematically in Fig.~\ref{fig:CRAD}. The detector consists of a cryogenic vessel with a vacuum 
thermal insulation that can hold up to 150 l of liquid Ar (200 kg). The sensitive volume is limited by the height 
of the bottom PMT system as well as the thickness of side PMTs and is close to 50 l (50 cm in height and 36 cm in diameter). 
Primary electrons, produced by recoil nuclei after WIMP or neutrino scattering, drift in the liquid towards the surface 
in the electric field of 2 kV/cm. At the liquid surface electrons are emitted in the gas phase where they produce signal 
in two stages: first by electroluminescence in the high electric field above the liquid and second by avalanche multiplication 
in a double-THGEM or in a hybrid 2THGEM/GEM cascade.

Proportional scintillations in the VUV region from an electroluminescence (EL) gap (S2 signal) are detected by the ring of side PMT 
and by the bottom PMT system. To provide sensitivity in the VUV range all PMTs have a wavelength shifter (tetraphenyl-butadiene, TPB) 
deposited either directly on the PMT window or on a thin transparent plastic foil in front of them. The bottom PMTs are also intended 
for the detection of primary scintillations in the liquid (S1 signal).

For the bottom PMT system an R11065 Hamamatsu 3-inch PMT was chosen based on the experience of WArP and DarkSide experiments ~\cite{PMT-WARP, PMT-DarkSide}. 
This PMT has a bialkali photocathode adapted for operation down to liquid Ar temperature (87 K). The quantum efficiency (QE) 
reported by the manufacturer is not less than 25\% at 420 nm. The
authors of  ~\cite{PMT-WARP} report, however, the value of 33\% QE at 420 nm. For the side PMT system the Hamamatsu R6041-506MOD device was chosen. 
This is a compact 2-inch PMT with a metal-channel dynode structure and a bialkali photocathode adapted for
operation in liquid Ar (87 K) with 25\% QE at 420 nm ~\cite{R6041specs}. This tube has 57 mm diameter and 43.5 mm height, which allows to keep 
the diameter of the sensitive volume not less than 36 cm.

\begin{figure}[htbp]
\centering
\includegraphics[width=0.7\textwidth,viewport=1 1 700 500,clip]{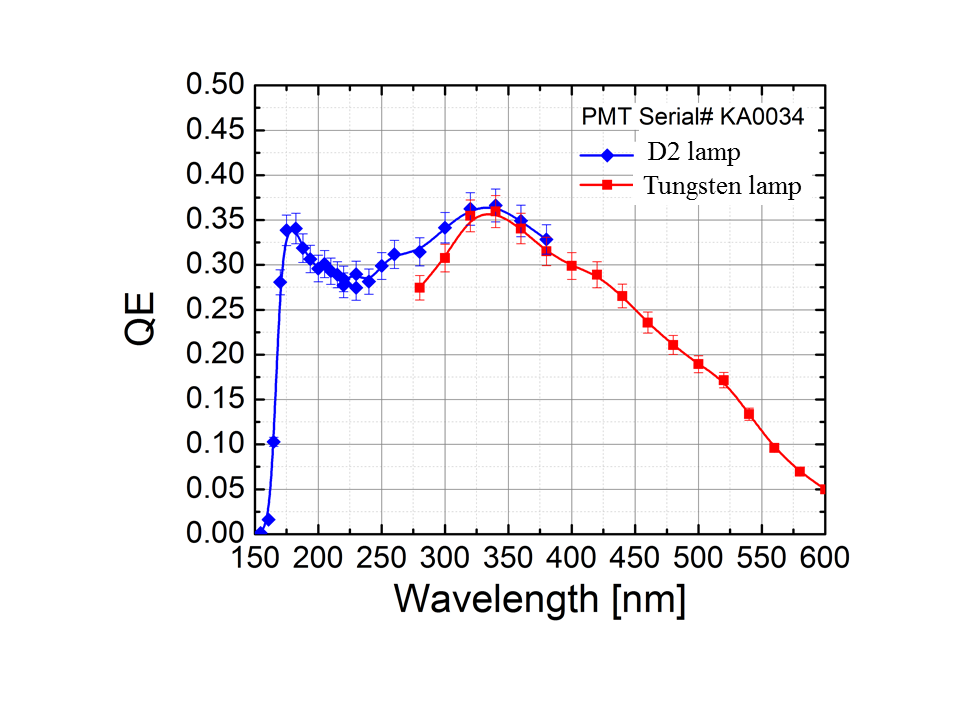}
\caption{Quantum efficiency of R11410 PMT. Result of the measurement from ~\cite{Lyashenko}.}
\label{fig:QER11410}
\end{figure}

The TPB layer considerably reduces a photon collection efficiency, by a factor reaching 20, due to internal reflection 
and conversion efficiency losses in the wavelength shifter (WLS) and because of the absence of optical contact between the WLS 
and the PMT window. Thus, it looks attractive to shift the VUV emission of Ar to a longer wavelength directly in the detection medium without the WLS. 
It is known that in liquid Ar such a VUV-to-UV conversion can be performed by doping with Xe, 
at a content of 10-1000 ppm (see ~\cite{doping_paper} and references therein). 
For this case we propose here to use a 3-inch R11410 PMT in liquid Ar conditions, 
originally designed for use in liquid Xe ~\cite{Xe-PMT1,Xe-PMT2,Xe-PMT3,Xe-PMT4}. 
This PMT has an enhanced sensitivity to UV photons emitted by Xe, with a quantum efficiency close to 30\% at 175 nm
(Fig.~\ref{fig:QER11410})~\cite{Lyashenko}). 
On the other hand, it has the same dynode system, the same body and the similar bialkali photocathode as those of R11065, 
which give hope for a successful operation at liquid Ar temperature.   

Another motivation for testing an R11410 PMT in liquid Ar conditions comes from the idea of the "universal argon-xenon detector". 
In this idea, the detector designed for operation in liquid Xe, in particular that of the RED experiment ~\cite{Xe-PMT4}, 
can also be operated with liquid Ar using the same PMT system, thus diversifying the target material.  

\section{Experimental set-up and conditions of the measurements}

\begin{figure}[htbp]
\centering
\includegraphics[width=0.7\textwidth,viewport=1 1 700 500,clip]{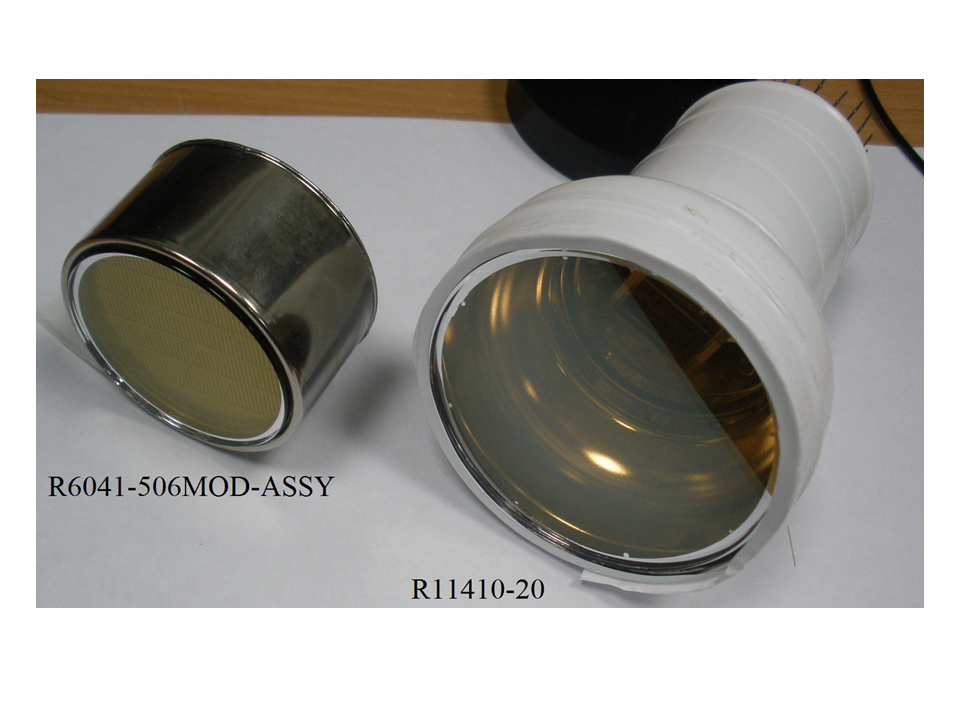}
\caption{Photo of R6041-506MOD and R11410-20 photomultipliers.}
\label{fig:PMTs}
\end{figure}

\begin{figure}[htbp]
\centering
\includegraphics[width=0.7\textwidth,viewport=1 1 700 500,clip]{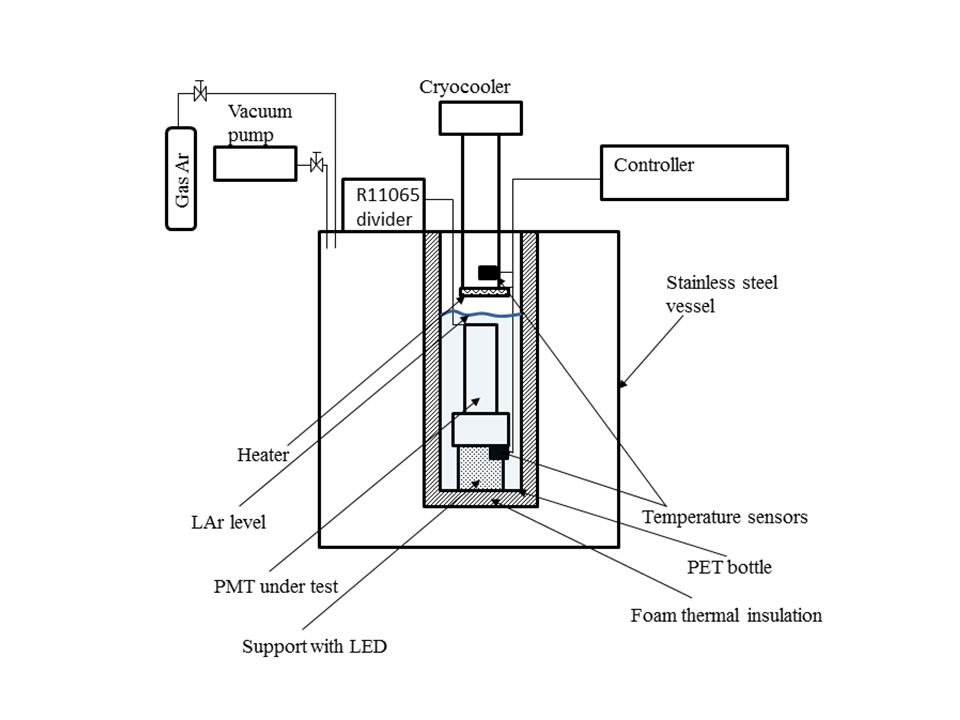}
\caption{Schematic set-up for the PMT tests.}
\label{fig:set-up}
\end{figure}

In our previous work  excellent performance of the R11065-10 and R11065-MOD PTMs
in liquid Ar was demonstrated ~\cite{CryoPMT1} and these results are presented here for completeness.  The total number of compact 
R6041-506MOD PMTs purchased from Hamamatsu Photonics  is 21 pieces. The results of the performance  of four PMTs 
were reported in ~\cite{CryoPMT1}. In the present  work the rest 17 PMTs are tested.  Seven 3-inch R11410-20 PMTs are tested 
in the present work and for the first time their successful operation in liquid Ar is demonstrated.  
The image of  R6041-506MOD and R11410-20 PMTs is shown in Fig.~\ref{fig:PMTs}. The set-up for testing of PMT immersed 
in liquid Ar is presented in Fig.~\ref{fig:set-up}. For the tests of PMT we used a stainless steel vessel 30 cm in diameter and 
40 cm high with a gas-tight top cover. The vessel can keep up to 3 atm of overpressure. The top cover contains 
three metal-glass feedthroughs that were used to provide high voltage to the PMT under test, signal from the pulse 
generator to the light emission diode (LED), signals from the temperature sensors to the Lake Shore 336 temperature controller 
and voltage to the heater from the same controller. Cryocooler Cryodyne 350CP by CTI-Cryogenics was used 
for cooling and liquefaction of Ar. The PMT under test was installed inside a thermal insulating cylindrical case 
10 cm in diameter and 35 cm high. The case has a plastic PET (polyethylene terephthalate) bottle 8 cm in diameter and
30 cm high firmly attached inside. At the bottom the bottle has special support containing the LED for PMT cathode illumination.

All three types of PMTs,  R11065, R11410-20  and R6041-506MOD were powered by positive voltage from N1470 CAEN high voltage 
power supply. A signal from the PMT anode was amplified by a factor of 10 or 100 with a CAEN N979 fast linear amplifier module. 
Then amplified signal was connected to a Tektronix TDS5034B oscilloscope  and to a discriminator CAEN N841 for counting rate 
measurements by a scaler CAEN N1145. The oscilloscope was used for the measurements of waveforms and pulse height spectra. 
Signals were obtained from spontaneous emission of single electrons from photocathode or by illumination with the LED. 
The LED was powered from a pulse generator, and its light intensity could be tuned by changing a pulse height and base line 
from the generator. The rate with LED did not exceed 5 kHz and the frequency of the pulse generator was 50 kHz.

\section{Results of the measurements}
\subsection{R11065-10, R11065-MOD}

\begin{figure}[htbp]
\centering
\includegraphics[width=0.7\textwidth,viewport=1 1 700 500,clip]{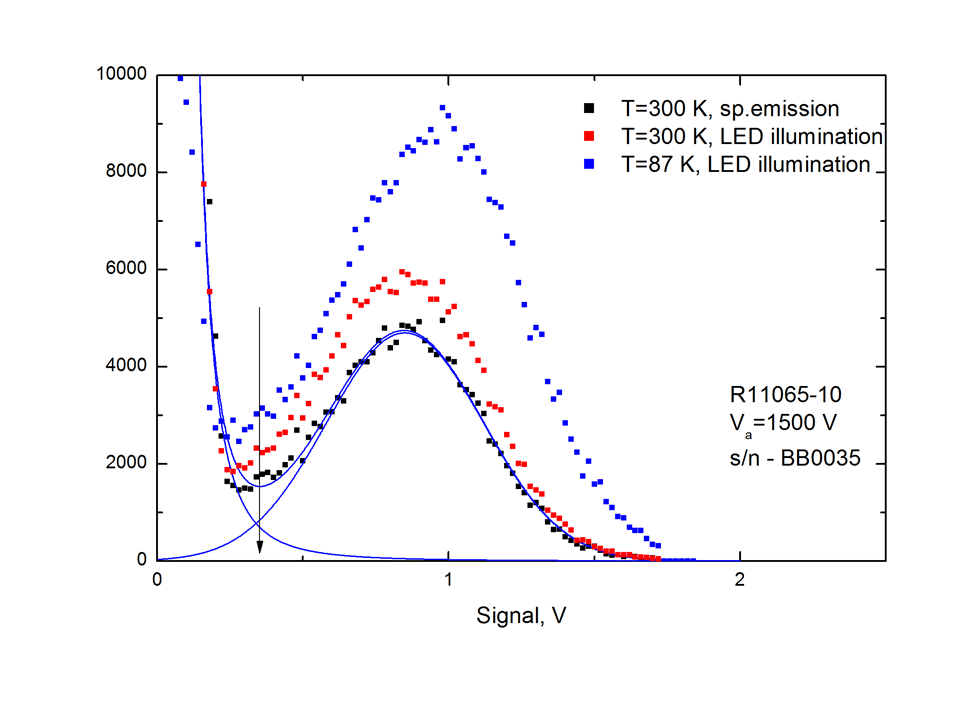}
\caption{Pulse height spectra of SER obtained from spontaneous emission (black points) and with LED illumination 
at room temperature (red points) and at 87$^{\circ}$K (blue points). Vertical arrow points to the threshold for calculation 
of the relative single electron efficiency. Anode voltage is 1500V and the  gain is ~10$^7$.
}
\label{fig:R11065spectrum}
\end{figure}

\begin{figure}[htbp]
\centering
\includegraphics[width=0.7\textwidth,viewport=1 1 700 500,clip]{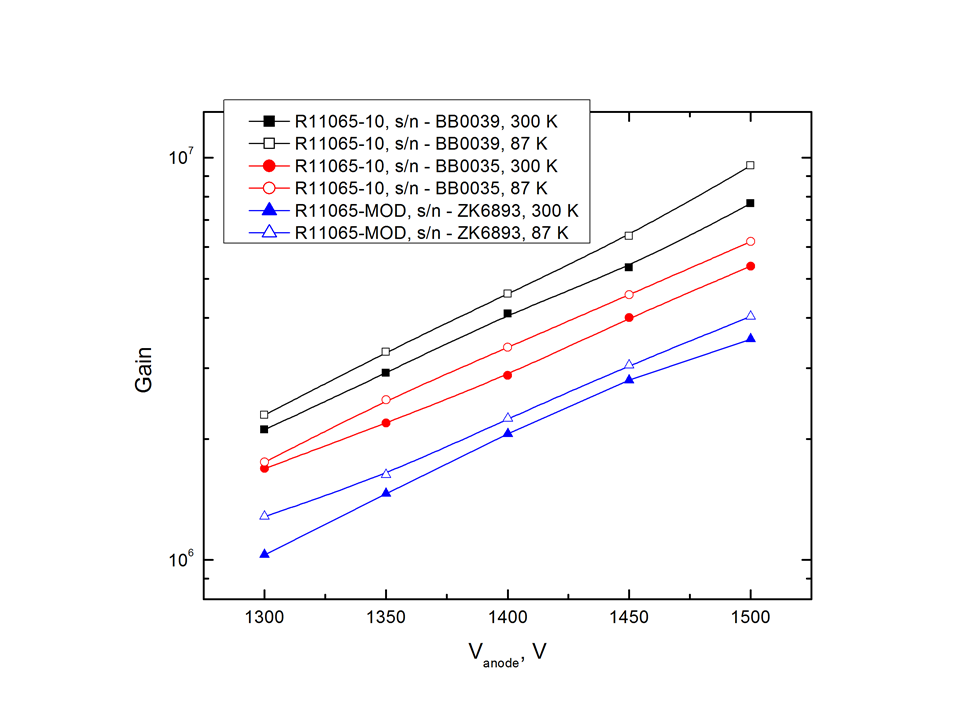}
\caption{Gain as a function of anode voltage for two R11065-10 (one with the highest maximum gain and one with the lowest maximum gain) 
and one R11065-MOD PMT. The results for different PMTs are marked with different colors. The data corresponding to liquid Ar temperature 
are marked with open points.}
\label{fig:R11065gain}
\end{figure}

\begin{figure}[htbp]
\centering
\includegraphics[width=0.7\textwidth,viewport=1 1 700 500,clip]{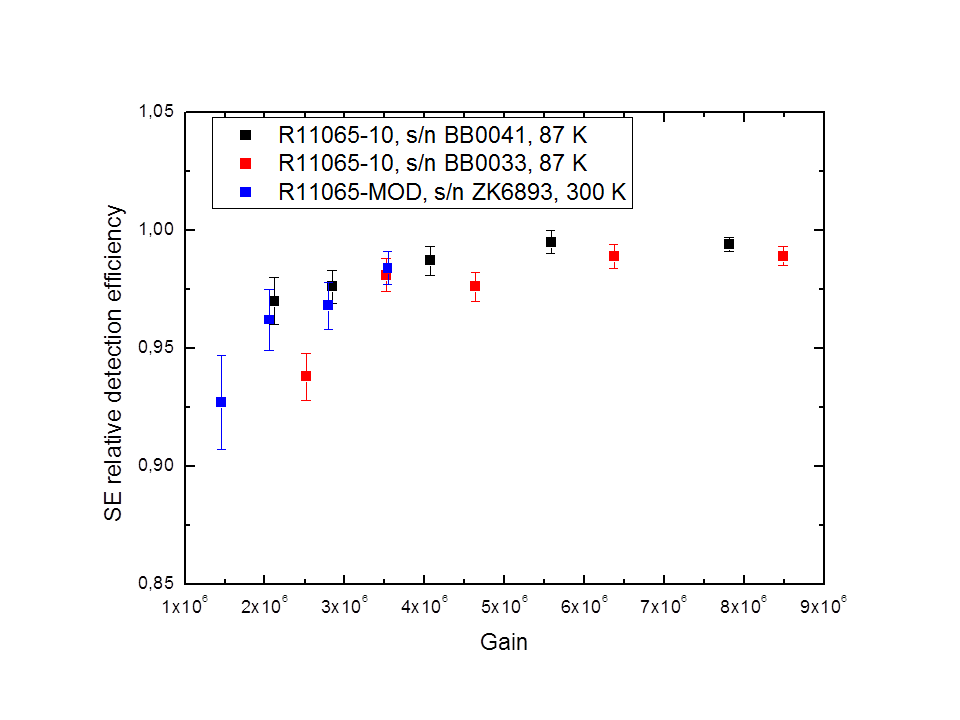}
\caption{Relative detection efficiency of single electrons as a function of PMT gain. The data are shown for three PMTs 
with the lowest, medium and the highest maximum gain.}
\label{fig:R11065eff}
\end{figure}

Single electron response (SER) of an R11065 tube is rather fast, FWHM of the signal is ~7 ns and full width at 0.1 maximum is around 12 ns.
The gain can reach values close to 10$^7$ and the single electron spectrum is very well separated from the noise (see Fig.~\ref{fig:R11065spectrum}). 
The gain of the PMT was calculated from the measurements of the pulse height spectra of SER. At first the waveform was measured for each  PMT: by integrating it and
expressing the total charge in electrons, the relationship between the full charge and the pulse height was calculated. Then the pulse height spectra were measured and the
most probable pulse height was found for each spectrum by fitting with the Gaussian distribution. The pulse height distribution of SER in PMT in the
absence of external noise sources is
mainly determined by multiplication statistics of the first dynode that is well described by the Poisson distribution. The Poisson distribution with
the average value from 7 to 9 (see ~\cite{CryoPMT1}) is very close to that of  the Gaussian. Moreover, the rest dynodes and electronic noise smear the final 
distribution following in average the Gaussian law.  
A gain as a function of anode voltage is shown in Fig.~\ref{fig:R11065gain} for three typical photomultipliers. A relative single electron efficiency 
(fraction of the SE response spectrum above threshold) reaches 95\% at the gain below 5x10$^6$ for all photomultipliers (Fig.~\ref{fig:R11065eff}).

\subsection{R6041-506MOD}

\begin{figure}[htbp]
\centering
\includegraphics[width=0.7\textwidth,viewport=1 1 700 450,clip]{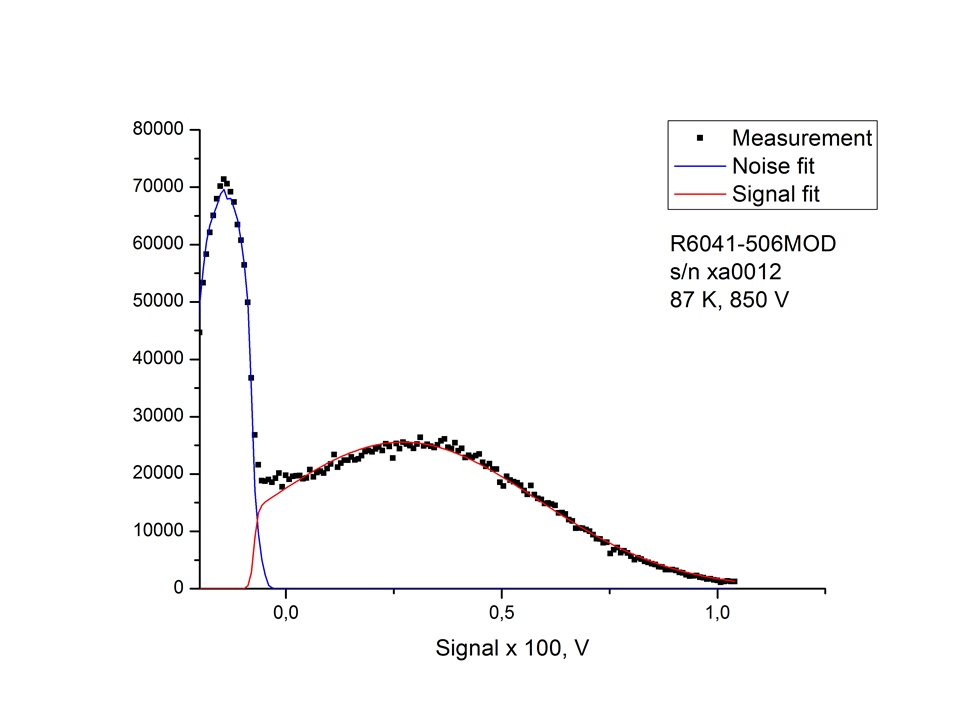}
\caption{Typical pulse height spectrum of SE response for a R6041-506MOD tube from a group of the PMTs with higher gain.}
\label{fig:R6041spectrum}
\end{figure}

\begin{figure}[htbp]
\centering
\includegraphics[width=1.0\textwidth,viewport=1 100 700 500,clip]{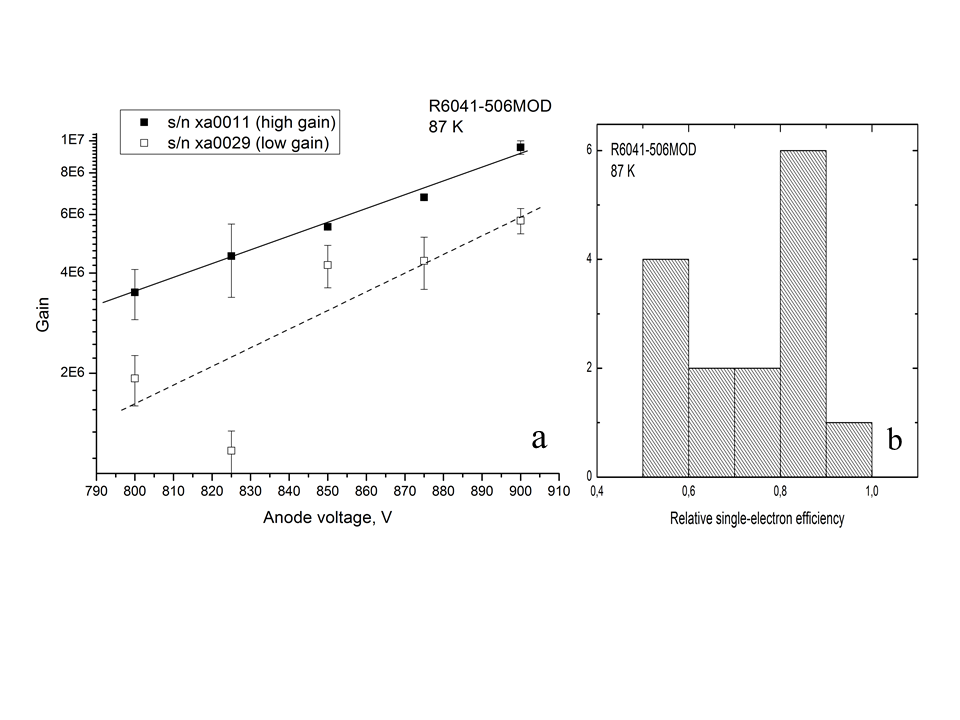}
\caption{a) Gain as a function of anode voltage for two typical R6041-506MOD tubes, one from the group with a higher gain 
and another from the group with a lower gain. 
b) Distribution of PMT over the relative SE efficiency. Only eight tubes show relative efficiency exceeding 75\%.  }
\label{fig:R6041gaineff}
\end{figure}

This PMT is about twice slower than R11065: FWHM of the waveform is 14 ns and the width at 0.1 of maximum is about 20 ns. 
As R6041-506MOD is a slower PMT as compared to R11065, it has an approximately twice lower average SE signal at the same gain. 
A typical SE response spectrum is shown in Fig.~\ref{fig:R6041spectrum}. As in the case of R11065, linear dependence between the pulse height and total
charge was used to calculate the gain. Since there was significant asymmetry of the SER spectrum,
the Poisson distribution convolved with the Gaussian function was used for the signal spectrum
approximation. The convolution allowed to account for the distribution smear due to the second- and higher-order
dynodes and external noises. To account for the discriminator used as the trigger for the data
collection, the signal model curve was multiplied by a steplike function with a narrow slope. Also
the noise spectrum measured at zero voltage was added to the fitting curve of the SER spectrum.
This noise was mainly caused by the electronics and was normally stable during the measurement
runs and independent of anode voltage. The average pulse height was calculated from the fitted
signal curve without considering discriminator cut off.
The gain of R6041-506MOD can reach 10$^7$, however, these tubes have rather large spread of  maximum gain and relative SE efficiency. 
Only eight photomultipliers  with  relative SE efficiency higher than 75\% could be selected (Fig.~\ref{fig:R6041gaineff}). In total only 12 PMTs 
out of 21 were selected for future operation in the two-phase CRAD. 

\subsection{R11410-20}

\begin{figure}[htbp]
\centering
\includegraphics[width=0.7\textwidth,viewport=1 1 700 550,clip]{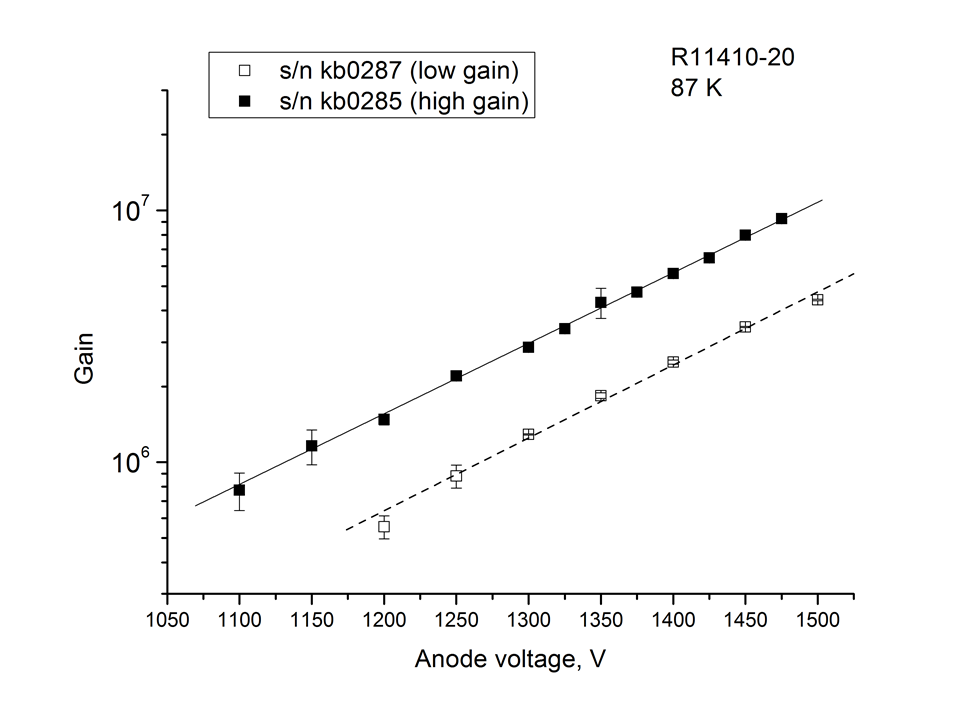}
\caption{Gain as a function of anode voltage for R11410-20 PMT with the highest and the lowest gain in the tested group.}
\label{fig:R11410gain}
\end{figure}

\begin{figure}[htbp]
\centering
\includegraphics[width=0.7\textwidth,viewport=1 50 700 500,clip]{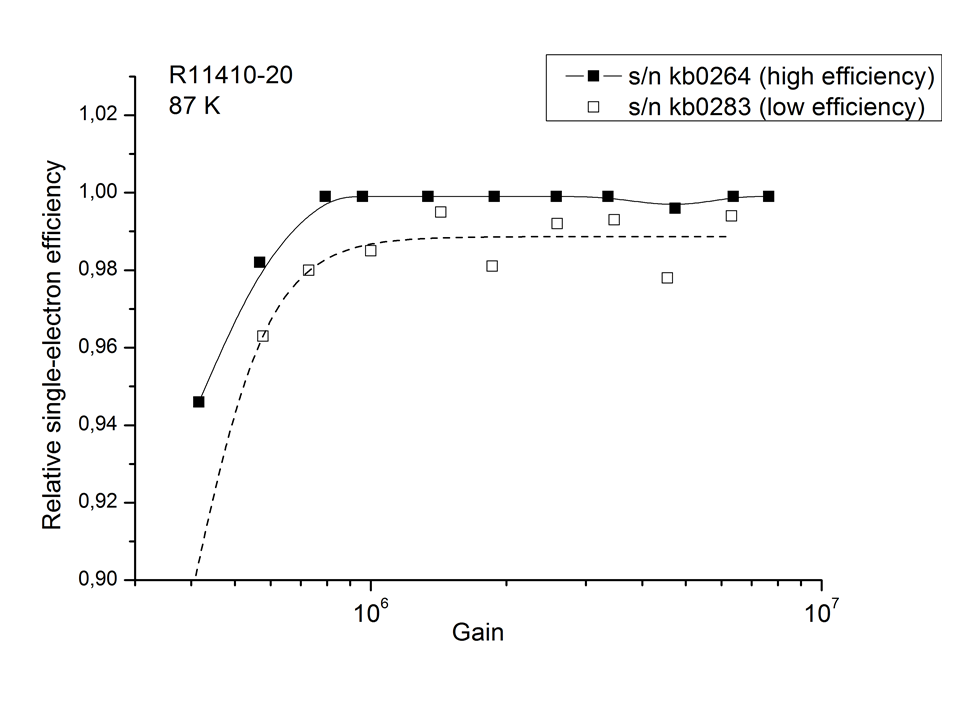}
\caption{Relative SE efficiency as a function of gain for two R11410-20 tubes, one with the highest 
and one with the lowest efficiency from the tested group.}
\label{fig:R11410eff}
\end{figure}

Performance of these photomultipliers is very similar to that of R11065 tubes. As we did not know whether these tubes could operate without discharges in liquid Ar
, the voltage was gradualy increased from 1000 V to 1500 V (recommended maximum voltage by the manufacturer). Four of the seven tested photomultipliers reached maximum
voltage without discharges operating in liquid Ar. The other three PMTs exhibit discharges at the anode voltage between 1300 V and 1375 V corresonding to the gain values around
3x10$^6$. After the discharges all parameters of the photomultipliers were recovered and anode voltage was kept below 1300 V. The pulse shape and the pulse height 
spectra of R11410-20 tubes look very much like that of R11065. Spread of gains, energy resolutions and relative efficiencies is rather small, 
unlike in the case of R6041-506MOD tubes. Typical gain-voltage curves and relative efficiency vs gain dependencies are shown 
in Fig.~\ref{fig:R11410gain} and Fig.~\ref{fig:R11410eff}. All seven tested photomultipliers demonstrated relative efficiency 
above 95\% in liquid Ar. 

\subsection{Long-term performance of the cryogenic PMTs.}

\begin{figure}[htbp]
\centering
\includegraphics[width=0.7\textwidth,viewport=100 1 700 550,clip]{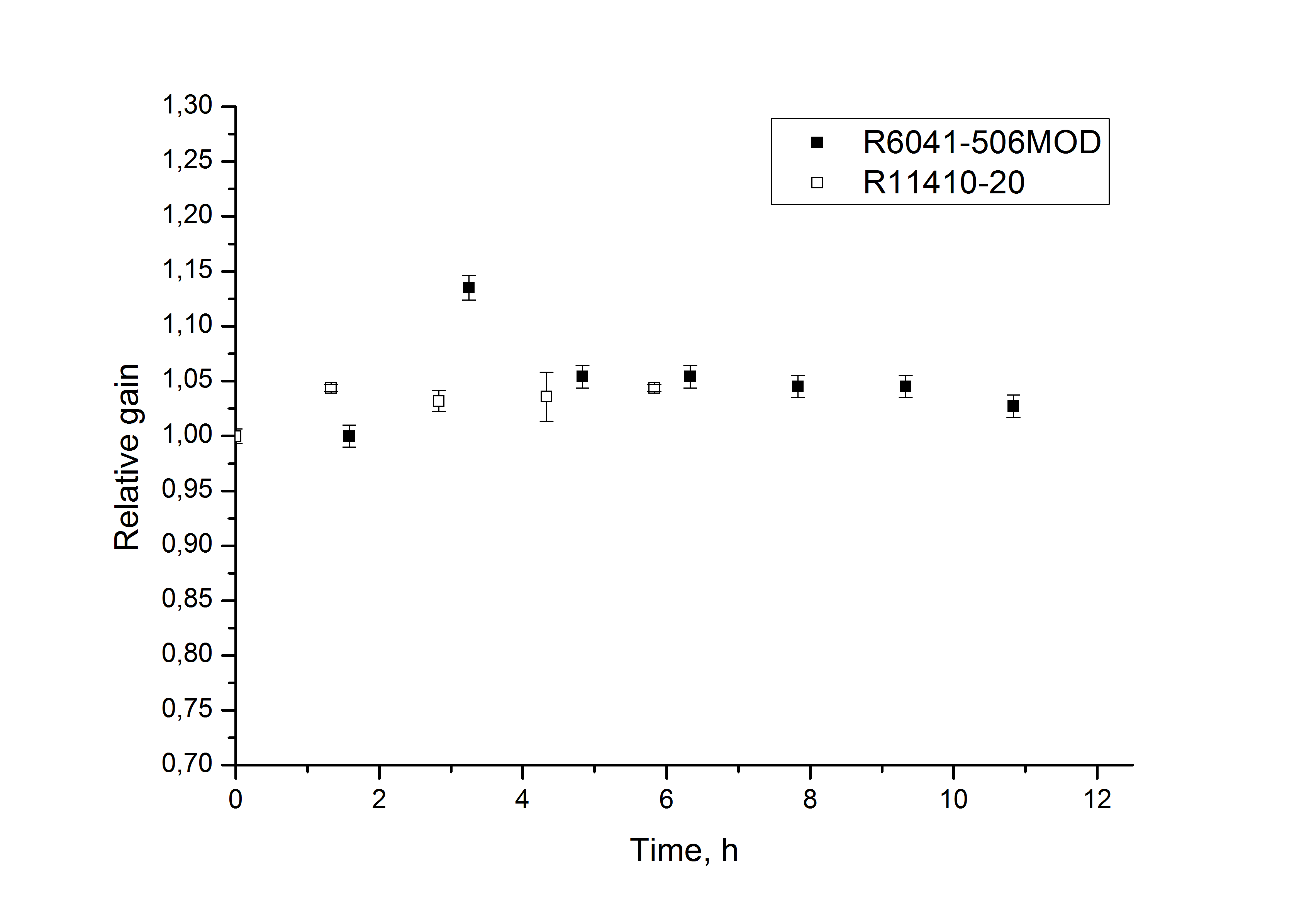}
\caption{Relative gain as a function of time for R6041-506MOD and R11410-20 photomultipliers.}
\label{fig:long_runs}
\end{figure}

The behavior of one R6041-506MOD and one R11410-20 PMTs immersed in liquid argon for the
period of several hours was studied. None of the gain, noise frequency or signal shape showed
significant change for both tubes during the runs. Figure ~\ref{fig:long_runs} shows gain vs time dependences for the
R6041-506MOD tube  and for R11410-20 photomultiplier. In both cases anode voltage was applied constantly during the whole run.

\section{Conclusions}

Operation of three types of the cryogenic PMT produced by Hamamatsu Photonics was tested in liquid Ar, namely 3-inch R11065-10 and R11065-MOD tubes,
2-inch compact R6041-506MOD tubes and 3-inch R11410-20 tubes sensitive to Xe UV emission. R11065 photomultipliers are designed for operation in liquid Ar.
They demonstrate excellent performance, stable operation with maximum gain reaching 10$^7$ and relative efficiency for single electrons higher than 95\%.
2-inch R6041-506MOD tubes are very attractive due to their compact design, however their gain and relative efficiency exhibit high spread and only 12
photomultipliers out of 21 were selected with relative SER efficiency higher than 75\% for operation in future two-phase CRAD. R11410-20 tubes were
designed for operation in liquid Xe and this work was the first demonstration of their operation in liquid Ar. These photomultipliers can safely operate 
at  87$^{\circ}$K with a maximum gain up to 10$^7$ and relative SER efficiency higher than 95\%.   

\acknowledgments

 This study was supported by Russian Science Foundation (project N 14-50-00080) and was done within the R\&D program of the DarkSide-20k experiment.

% We suggest to always provide author, title and journal data:
% in short all the informations that clearly identify a document.

\end{document}